\documentclass{iaus}
\usepackage{graphicx}

\title[SpS 7.~~Stellar and brown dwarf properties] 
{ Stellar and brown dwarf properties \\ from numerical simulations}

\author[Matthew R. Bate]   
{Matthew R. Bate
}

\affiliation{
School of Physics, University of Exeter \\ Stocker Road, Exeter EX4 4QL, United Kingdom \\ email: {\tt mbate@astro.ex.ac.uk} \\
}

\pubyear{2008}
\volume{xxx}  
\pagerange{119--126}
\jname{Young Stars, Brown Dwarfs, and Protoplanetary Disks}
\editors{Jane Greg—rio-Hetem \& Silvia Alencar, eds.}
\begin{document}

\maketitle

\begin{abstract}
We review the statistical properties of stars and brown dwarfs obtained from the first
hydrodynamical simulation of star cluster formation to produce more than a 
thousand stars and brown dwarfs while simultaneously resolving the lowest mass 
brown dwarfs (those with masses set by the opacity limit for fragmentation), binaries 
with separations down to $\sim 1$ AU, and discs with radii greater than $\sim 10$ AU.  
In particular, we present the eccentricity distribution of the calculation's very-low-mass
and brown dwarf binaries which has not been previously published.
\end{abstract}


Recently, \cite[Bate (2009a)]{Bate2009a} published hydrodynamical simulations 
of the collapse and fragmentation of a 500 M$_\odot$ molecular cloud with 
a diameter of 0.8~pc, and a mean thermal Jeans mass of 1 M$_\odot$.  
Although initially spherical, the cloud was seeded with a supersonic
divergence-free `turbulent' velocity field with a power spectrum $P(k) \propto k^{-4}$.  
The turbulent energy was initially set equal in magnitude to the gravitational 
energy of the cloud and decayed during the calculations.  A barotropic equation of 
state that mimics the heating of collapsing gas once it becomes optically thick to its own 
radiation was used, thus capturing the opacity limit for fragmentation.  Once the central density of a 
protostar exceeded $10^{-10}$~g~cm$^{-3}$, the object was replaced by a sink particle
that accreted any gas approaching within a specified accretion radius.  These objects began
with masses of a few Jupiter masses, and subsequently accreted to become more massive brown
dwarfs or stars.  The same calculation was performed twice with two different accretion
radii (5 AU and 0.5 AU).  The former was run until 1.5 initial cloud free-fall
times (285,000 yrs) by which time 1254 stars and brown dwarfs had formed. 
Due to shorter timesteps, the latter was only followed to 1.04 free-fall times, producing
258 objects.  Comparison allowed the effect of different accretion radii on the 
statistical properties of the stars to be assessed.  A brief summary of the statistical
properties of the stars and brown dwarfs produced by the calculations is given below.
In most cases, more detail can be found in \cite[Bate (2009a)]{Bate2009a}.

\begin{figure}
  \begin{center}
  \caption{Masses and orbital properties of the 16 binary VLM objects produced by the star cluster formation calculation (\cite[Bate 2009a]{Zinner98}) using sink particle accretion radii of only 0.5 AU. The shaded histogram contains those binaries with semi-major axes less than 30 AU.  These close systems are more likely to survive to join the field population and tend to have low eccentricities.}
  \label{bate_fig1}
 {\scriptsize
  \begin{tabular}{|c|c|c|c|}\hline 
M$_1$	& M$_2$	&	Semi-major axis	& Eccentricity \\
M$_\odot$ & M$_\odot$ & AU				& \\ \hline
0.096	& 0.065	& 0.52	& 0.365  \\
0.095	& 0.082	& 0.95	& 0.045 \\
0.044	& 0.014	& 1.13	& 0.493 \\
0.017	& 0.021	& 1.75	& 0.419 \\
0.041	& 0.021	& 2.84	& 0.233 \\
0.067	& 0.040	& 3.77	& 0.514 \\
0.076	& 0.057	& 4.05	& 0.097 \\
0.047	& 0.029	& 10.5	& 0.075 \\
0.056	& 0.051	& 21.3	& 0.036 \\
0.047	& 0.041	& 47.1	& 0.511 \\
0.025	& 0.010	& 49.6	& 0.261 \\
0.079	& 0.056	& 61.9	& 0.546 \\
0.057	& 0.023	& 93.6	& 0.805 \\
0.061	& 0.015	& 226	& 0.566 \\
0.080	& 0.026	& 577	& 0.324 \\
0.099	& 0.051	& 989	& 0.861 \\ \hline
  \end{tabular}
  }
  \hspace{0.5cm}
 \includegraphics[width=2.4in]{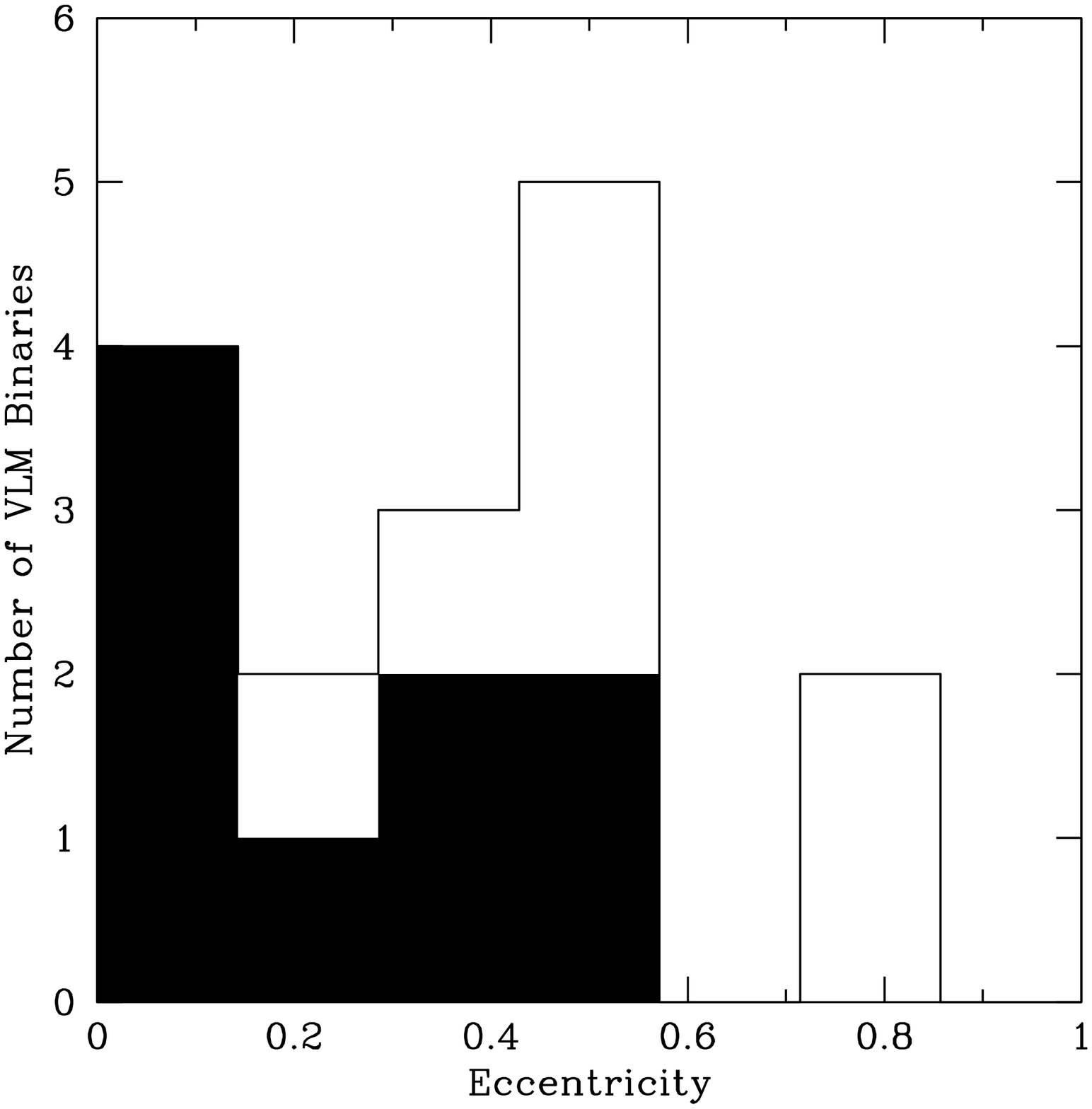} 
 \end{center}
\end{figure}


{\underline{\it Multiplicity}}. The calculations produced stellar populations in which
multiplicity is a strongly increasing function of primary mass with values in good 
agreement with observational surveys.  In particular, the frequency of 
very-low-mass (VLM) binaries with primary
masses of $0.03-0.10$~M$_\odot$ is found to be $19\pm 5$\% for the 0.5 AU accretion
radius calculation, in good agreement with observations.  We also note that 
multiplicity is predicted to continue decreasing with primary mass throughout the
brown dwarf regime such that the predicted multiplicity of $0.01-0.03$~M$_\odot$
brown dwarfs is less than 7\%.

{\underline{\it Binary and multiple separations}}. The separations of binary and 
higher-order multiple systems are also found to depend on primary mass,
with stellar systems having a median separation of 26 AU and VLM systems having
a median separation of 10 AU.  This tendency for lower mass binaries to have smaller
separations is in agreement with observations, although observations indicate than
VLM systems are even tighter than those produced in the calculations.  However,
we also note that the more accurate 0.5 AU accretion radius simulation produces 
tighter binaries than the 5 AU accretion radius simulation and also that the
separations of VLM binaries decrease during the simulation.

{\underline{\it Binary eccentricities}}. The eccentricity distribution of binaries is 
very sensitive to the accretion radius.  The calculation using 5 AU 
accretion radii produced a severe excess of high eccentricity ($e>0.7$) close 
($<10$ AU) systems.  This was corrected when accretion
radii of 0.5 AU were used, and the mean eccentricity is then in good agreement with
observations.  \cite[Bate (2009a)]{Bate2009a} does not differentiate
between stars and brown dwarfs when discussing eccentricities.  However, at this
meeting, preliminary results for the eccentricity distribution of VLM binaries were 
presented by Trent Dupuy (see the contribution by Dupuy, this volume).  
He finds VLM binaries tend to have low eccentricities (usually less than 0.5).  In 
Fig.\,\ref{bate_fig1}, we present the properties of the 16 VLM binaries produced by the
0.5 AU accretion radius calculation of \cite[Bate (2009a)]{Bate2009a}.  In
agreement with the new observational results, the simulated VLM binaries tend to
have low eccentricities, particularly those with separations less than 30 AU which is
typical of the field population.

{\underline{\it Model deficiencies and conclusions}}. As discussed above, many of the 
stellar properties obtained from the simulations are in good agreement with observations.  
The implication is that the properties of binary and multiple systems are primarily 
determined by the combination of gravity and gas dynamics (i.e.\ dissipative gravitational 
dynamics).  Other processes such as radiation transport and magnetic fields 
are not required to explain the origin of these properties.  However, the 
hydrodynamical simulations do have two major deficiencies: they give a ratio of brown dwarfs 
to stars much higher than is observed (roughly 2:1) and there is a
deficit of unequal-mass solar-type binaries. Recently, \cite[Bate (2009b)]{Bate2009b} 
showed that including radiative feedback may correct the ratio of brown 
dwarfs to stars, but the origin of the unequal-mass solar-type binary deficit is still unknown.




\end{document}